\documentclass[manuscript]{aastex}

\slugcomment{Submitted to New Astronomy}

\shorttitle{Variability of Mrk 501}
\shortauthors{Gupta et al.}

\begin{document}


\title{Multi-color Optical Variability of the TeV Blazar 
Mrk 501 in the Low-State}


\author{A. C. Gupta\altaffilmark{1}, W. G. Deng\altaffilmark{2,1},
U. C. Joshi\altaffilmark{3}, J. M. Bai\altaffilmark{1} 
and M. G. Lee\altaffilmark{4}}


\email{acgupta30@gmail.com, Phone No. +86 15920411403, Fax No. +86 871 3920599}




\altaffiltext{1}{National Astronomical Observatories / Yunnan Observatory, Chinese 
Academy of Sciences, P.O. Box 110, Kunming, Yunnan 650011, China.}
\altaffiltext{2}{Department of Physics, Yunnan University, Kunming, Yunnan 650091, China.} 
\altaffiltext{3}{Astronomy and Astrophysics Division, Physical Research Laboratory, 
Navrangpura, Ahmedabad - 380 009, India.}
\altaffiltext{4}{Astronomy Program, Seoul National University, Seoul 151-742, South Korea.}


\begin{abstract}

We report results based on the monitoring of the BL Lac object Mrk 501
in the optical (B, V and R) passbands from March to May 2000. Observations
spread over 12 nights were carried out using 1.2 meter Mount Abu Telescope,
India and 61 cm Telescope at Sobaeksan Astronomy Observatory, South Korea. The
aim is to study the intra-day variability (IDV), short term variability
and color variability in the low state of the source. We have detected flux 
variation of 0.05 mag in the R-band in time scale of 15 min in one night. 
In the B and V passbands, we have less data points and it is difficult to infer 
any IDVs. Short term flux variations are also observed in the V and R bands during 
the observing run. No significant variation in color (B$-$R) has been 
detected but (V$-$R) shows variation during the present observing run. 

Assuming the shortest observed time scale of variability (15 min) to
represent the disk instability or pulsation at a distance of 
5 Schwarschild radii from the black hole (BH), mass of the central BH is 
estimated $\sim$ 1.20 $\times$ 10$^{8} M_{\odot}$.

\end{abstract}






\section{Introduction}

Blazars constitute a class of radio-loud active galactic nuclei (AGNs) consisting of 
BL Lacertae objects (BL Lacs) and flat spectrum radio quasars (FSRQs). In the 
unified schemes of AGNs, blazars are believed to be the beamed counterparts of the 
FR I/FR II radio galaxies and many of their properties can be well understood in 
such a scenario (e.g. Browne 1983; Antonucci 1993; Urry \& Padovani 1995). According 
to orientation based unification scheme for radio-loud AGNs, blazar's jet make angles 
$\leq$ 20$^{\circ}$ or so to the line of sight (Urry \& Padovani 1995). Significant 
flux variations in the whole range of electromagnetic spectrum on time scale of less 
than a day to several years are common in blazars. Their radiation at all wavelengths 
is predominantly non thermal and strongly polarized ($> 3\%$) in radio to optical
bands.  

Blazar variability can be broadly divided into 3 classes viz. micro variability or
intra-night variability or intra-day variability, short term outbursts and long term 
trends. Variations in flux of a few tenth of a magnitude in a time scale of tens 
of minutes to a few hours (less than a days) is often known as intra-day variability 
(IDV) (Wagner \& Witzel 1995). Short term outbursts can range from weeks to months and 
long term trends can have time scales of months to several years.

Study of optical and near-IR variability of blazars on diverse time scales was carried out 
by several groups (Miller \& McGimsey 1978; Xie et al. 1988; Webb et al. 1988; Kidger \& de 
Diego 1990; Carini 1990; Carini et al. 1992; Heidt \& Wagner 1996; Bai et al. 1998, 1999; 
Fan \& Lin 1999; Ghosh et al. 2000; Gupta et al. 2002, 2004; Stalin et al. 2005; and references 
therein). The first convincing evidence of optical IDV was found in BL Lacertae by Miller et al. 
(1989). The typical optical IDV found in most of the blazars is $\sim$ 0.01 mag hr$^{-1}$. 
Carini (1990) observed a sample of 20 blazars and found most of them showing detectable 
inter-night variations. He also noted that in observing runs of more than 8 hours, the probability 
of detecting significant IDV was $\sim$ 80\%. Subsequently, Heidt \& Wagner (1996) observed a sample 
of 34 radio selected BL Lacs and detected IDV in 28 (i.e. 82\%) sources out of which 75\%  sources 
showed significant variation with in $<$ 6 hours which further supports the finding of high frequency 
of occurrence of IDV in blazars. In our recent paper (Gupta \& Joshi, 2005), we studied the frequency 
of occurrence of optical IDV in different classes of AGNs and found about 10\% (18/174) radio-quiet 
AGNs, 35$-$40\% (41/115) radio-loud AGNs (excluding blazars) and 70\% (79/113) blazars to show IDV 
(in fact $\sim$ 80$-$85\% blazars show IDV, if observed for more than 6 hours).

Mrk 501 is one of the most interesting and nearby BL Lac object at z = 0.034
which makes it the second closest known BL Lac object after Mrk 421 (z =
0.031).  Mrk 501 has been studied for variability in all regions of the
electromagnetic  spectrum (e.g. Fan \& Lin 1999, Kataoka et al. 1999,
Sambruna et al. 2000, Ghosh et al.  2000, Xue \& Cui 2005, Gliozzi, et al.
2006, and references therein). In optical region  this source has been studied
by several groups e.g Stickel et al. (1993) have reported  variation of 1.3
magnitude in luminosity; Heidt \& Wagner (1996) have reported $\sim$ 32\% 
variation in flux in less than two weeks time; Ghosh et al. (2000) have made 10
nights  observations on Mrk 501 during March to June, 1997 to search for
rapid variability  and reported rapid variability in 7 nights;
and recently, Fan et al. (2004) observed  Mrk 501 for two hours in one night
in October 2000 and found no significant variation.

The variability in blazars is mainly attributed to either due to the disc instability 
or the activity in the jet. IDV reported in the flaring and high state is generally 
attributed to the shock moving down the inhomogeneous medium in the jet. But small 
amplitude variation in the low state of blazar could be due to instability in the 
accretion disk (Witta et al. 1991, 1992; Chakrabarti \& Wiita 1993; Mangalam \& 
Wiita 1993). Recently, it is noticed that, in the low luminosity AGNs, accretion disk is
radiatively inefficient (Chiaberge et al. 2006; Chiaberge \& Macchetto 2006; Macchetto
\& Chiaberge 2007; Capetti et al. 2007). So, there will be an alternative way to 
explain the IDV in the low-state of blazars, in which a weak jet emission will   
be responsible for the IDV. So far there is no clear answer to this. There is a need 
to study IDV and short term variability in the optical bands with a good sampling in time 
when the source is in low state. In view of this we pursued the present study to address 
both the IDV and short term variability on Mrk 501. 

This paper is structured as follows: section 2 describes photometric observations 
and data reductions, section 3 present the results, and 
discussions and conclusions are presented in section 4. 

\section{Observations and Data Reduction}

Observations were made from two places: Mount Abu Observatory of Physical Research 
Laboratory, India and Sobaeksan Optical Astronomy Observatory, South Korea. \\
{\bf Observations from Mount Abu Observatory:} \\
Photometric monitoring of Mrk 501 was done in B Johnson and  R Cousins passbands 
using a thinned back illuminated Tektronix 1K $\times$ 1K  CCD detector at f/13 Cassegrain 
focus of the 1.2 meter telescope at Gurushikhar, Mount Abu, India. To improve the S/N ratio, 
2 $\times$ 2 on chip binning was done resulting in pixel size of 0.634 arcsec in both 
the dimension of the sky. The entire CCD chip covers $\sim$ 5.4 $\times$ 5.4 arcmin$^{2}$ 
of the sky. The readout noise and gain of the CCD chip were 4 electrons and 10 electrons/ADU
respectively. Exposure time for B and R passband were 300s and 150s respectively. 
The average seeing during the observing run remained at 1.5 arcsec. 
For flat field correction, twilight flats and bias frames were taken every night.

For each night median bias and median flat frames were constructed which 
were used in image processing. Image processing (bias subtraction, flat-fielding and cosmic 
rays removal), photometric reduction (getting instrumental magnitude of stars and Mrk 501) 
and calibration were done at Physical Research Laboratory, Ahmedabad, India. 
Image processing was done using standard routines in the IRAF package.
 
Star no. 1, 4 and 6 (see Villata et al., 1998)
 in the field of Mrk 501 were selected as comparison stars.
But, for the observing run  March 8-10, 2006, star no. 1 was saturated in 6 
 out of 9 R-band image frames and star no. 4 was found to be unstable, so, 
we used star no. 6 to calibrate the Mrk 501 data obtained in March 2000. 
Observations taken in April 2000 were calibrated using star no. 1. In April 2000 observations
star no. 4 and 6 were having poor S/N ratio compared to Mrk 501.  
There is advantage, sky variation (seeing) and cloud effects are effectively eliminated by
Instrumental magnitude of Mrk 501 and standard stars in the field of Mrk 501 were determined  
with the help of DAOPHOT II package 
(Stetson 1987, 1992) using concentric aperture photometric technique taking
apertures of radii 5.0, 7.0, 9.5 and 12.0 pixels. 
The data reduced with different  aperture radii were found in good agreement. However, it 
was noticed that the 
best signal to noise ratio was obtained with aperture radius of 
9.5 pixels (6 arcsec $\sim$ 4 $\times$ FWHM).  
Stars in different image frames of Mrk 501 field were cross matched by using DAOMATCH 
routine in DAOPHOT II package. \\
{\bf Observations from Sobaeksan Optical Astronomy Observatory:} \\
Photometric observations of Mrk 501 were also carried out in V and R passbands
 with the 61 cm 
Richey-Chretian telescope (f/13.5) and PM512 CCD Camera at the Sobaeksan Optical Astronomy 
Observatory, South Korea. The field of view of the CCD image is 4.3 $\times$ 4.3 arcmin$^{2}$ 
and its pixel scale is 0.5 arcsec/pixel. The gain is 9 electrons/ADU, and the readout noise 
is 10 electrons. The exposure time varies between 150s and 300s. Maximum exposure time is 
300s due to unstable tracking of the telescope. The CCD uses an LN2 cooling system. Bias 
images were taken at the beginning and the end of the observation. Sky flat images were 
taken at both dusk and dawn. 
The instrumental magnitudes were obtained using the aperture photometry 
routine in IRAF/DAOPHOT package as discussed above. 
For aperture photometry, aperture radii were decided depending on the seeing, 
that is, 3 or 4 times of the FWHM.
In the image frames 3  comparison (standard) stars, no. 1, 4 and 6 were present. 
Data are calibrated 
using star no. 1. We did not use star no. 4 and 6 for calibration, since these stars 
were having poor S/N ratio compared to Mrk 501.

Observation log and the intra-day variability results are listed in Table 1. 
Photometric data in B, V and R passbands are given in Table 2, 3 and 4 respectively. 
Photometric software DAOPHOT does not give the actual internal error of brightness but 
it gives the photon noise. The internal photometric errors of brightness for each passband 
are estimated using artificial addstar experiment as described by Stetson (1987). We found 
the standard deviation ($\sigma$) = 0.015, 0.012 and 0.010 for B, V and R passbands respectively.

\section{Results}

\subsection {Flux and Color Variations}

\subsubsection {Intraday Variability}

Light curves of comparison star (instrumental mag.), Mrk 501 (instrumental mag.)
and Mrk 501 (standard mag.) in V and R passbands are plotted in Fig. 1 and Fig. 2. 
For plotting purpose, only the data for those nights are taken where the data points 
are more than seven. We report here objectively the presence or absence of IDV 
at a confidence level of more than 5$\sigma$.

{\bf B Passband:} The source has been observed in B passband on 3 nights 
but there are only a few data points (three) in each night and hence the 
data are not plotted. 
Measurements in B passband show variation to the extent of 0.013 mag on March 8, 0.080 
mag on March 9 and 0.015 mag on March 10. Looking at the data of March 9, we notice 
fluctuations in brightness of comparison star, which might be partly responsible for
the apparent variation of 0.015 mag seen in Mrk 501, so the 
variation detected is doubtful. Hence as per our criterion, there is no
significant IDV seen in B band data. 

{\bf V Passband:} Light curves in V passband show variation to the extent of 0.079
mag on March 30 (Fig. 1 (a)). Visual inspection of the figure indicates that the
the variation seen
in Mrk 501 might be due to the fluctuations in comparison star.
So, the variation seen may not be a genuine IDV. 
Variations are also noticed on other nights: 0.019 mag on March 31, 0.026 mag on May 1, 
0.035 mag 
on May 2 and 0.034 mag on May 3. However, as per our IDV detection criterion,
we conclude no detection of significant IDV during the observing run of 5 nights.

{\bf R Passband:} Light curves in R passbands show variation to the extent of 
0.004 mag on March 8, 0.027 mag on March 9, 0.009 mag on March 10,
0.022 mag on March 30, 0.029 mag on March 31, 0.147 mag on April 1, 0.053 mag
on April 5, 0.018 mag on April 6, 0.021 mag on May 1, 0.077 mag on May 2 and
0.039 mag on May 3. The data on April 1 are noisy, therefore
the variation of 0.147 mag seen in this night is not taken as genuine IDV.
Visual inspection of light curve on May 2 shows fluctuations in the magnitude of
comparison star and hence the apparent variation seen in Mrk 501 might be
partly due to this. Therefore, the variation seen is not considered as real.
So, out of 11 nights observations in R passband, IDV is seen 
only in 1 night, i.e April 5.

Variations, similar to that seen in the present work, have also
been  noted earlier by other researchers e.g. on one occasion Ghosh et
al. (2000) have reported variation of 0.13 mag in V band in less than 30
min while on the other occasion they report brightness variation of
0.04 mag in 29 minutes followed by steady brightness. According to
Ghosh et al. (2000), the source was not in active phase during their
observing run. Fan et al. (2004) have reported the results on Mrk 501
based on the continuous observing run of two hours in October 2000 in
R band and no significant variations were noticed. A blazar generally
show IDV of relatively large amplitude during its outburst  period
while it shows very small flux variation or no obvious variation
during the quiescent state. Thus during our observing run, Mrk 501
appears to be in quiescent state as the intraday brightness variations
detected are less than  6 to 7\%.

\subsubsection{Short Term Variability}

In Fig. 4 we have plotted all the data points in B, V and R passbands
against the time in JD. The inter-night brightness  variations are
distinctly visible in the figure; the maximum variation of the  source
in V passband being 0.19 mag (between its faintest level at 13.78 mag 
on JD 2451635.320 and the brightest level at 13.59 mag on JD
2451667.237) whereas  in R passband it has varied by 0.29 mag (between
its faintest level at 13.46 mag  on JD 2451613.505 and the brightest
level at 13.17 mag on JD 2451667.194).
Data of  April 1 being noisy, are not included in the short term 
variability analysis. 
 During the observing run of about 8 weeks, the source shows 
brightness variation of about $\sim$ 15 \% in V passband and $\sim$ 25 \%
in R passband. Heidt \& Wagner (1996) have reported $\sim$ 32\%
variation in flux in less than two weeks time which is larger than the
variation seen in the present work 
(the maximum brightness variation detected in the present work is about 25\% 
in about 2 months time).

\subsubsection{Color Variation}

In Fig. 4 panel (d) and (e), nightly  average color (B$_{ave}$ $-$ R$_{ave}$) 
and (V$_{ave}$ $-$ R$_{ave}$) respectively are plotted against the time (in JD).
For plotting purpose instead of taking the mean time (in JD) of the observing 
run, we have taken JD at 00h 00m 00s UT of that specific date. In Fig. 2, panel 
(d), there is no significant change in (B$_{ave}$ - R$_{ave}$) color but in panel 
(e) the variation in color (V$_{ave}$ $-$ R$_{ave}$) is apparent. The effect of 
variation in V and R magnitude (see panel (c) and (d)) has reflected in the 
(V$_{ave}$ - R$_{ave}$) color showing, the source has become  
0.19 mag bluer during the observing
campaign (between 0.53 at JD 2451635.50 and 0.33 at JD 2451669.50).  

\subsection{Variability time scale}

Fig. 2, panel (g) shows the light curve of Mrk 501 in R-band on April 5, 2000. 
During the period JD 2451640.412 to JD 2451640.421, the luminosity has decreased 
by 0.04 mag and then a rapid flare has occurred at epoch JD 2451640.421, reaching 
a brightness excursion of 0.05 mag at JD 2451640.431. Just after the flare, again 
there is a decrease in luminosity of 0.02 mag between JD 2451640.431 and 
JD 2451640.438. There after the source became steady
(see Fig 2 (g) top panel).

To search  for the 
time scale of variation quantitatively, we carried out Structure Function (SF) analysis 
on the R-band IDV of Fig. 2 (g) (top panel). 
SF is usually calculated twice by using an interpolation algorithm, first starting
from the beginning of the time series and proceeding forward and then starting from 
the end and proceeding backward. This will give two slightly different SF curves
but will provide a rough assessment of the errors due to the interpolation process   
(Wu et al. 2006).

The first order structure function is equivalent to the power spectrum
density (PSD) and is a powerful tool to search for periodicities
and time scales in a time series data (see e.g. Rutman 1978; Simonetti,
Cordes \& Heeschen 1985; Paltani et al. 1997). The
first order SF for a light curve having uniformly spaced data points
is defined as
\begin{eqnarray}
D_{a}^{1} (k) = \frac {1}{N_{a}^{1}(k)} \sum_{i=1}^{N} w(i) w(i+k)
[a(i+k) - a(i)]^{2}
\end{eqnarray}
where k is time lag, ${N_{a}^{1}(k)} = \Sigma w(i) w(i+k)$ and the weighting
factor w(i) is 1 if a measurement exists for the ith interval, 0 otherwise.
The squared uncertainties in the estimated SF is
\begin{eqnarray}
\sigma^{2}(k) = {\frac {8\sigma^{2}_{\delta f}}{N^{1} (k)}} D^{1}_{f} (k)
\end{eqnarray}
where $\sigma^{2}_{\delta f}$ is the measured noise variance.

Since the data sampling in our light curve is quasi-uniform, we used
interpolation method to determine the first order SF. For any time lag k,
the value of a(i+k) was calculated by linear interpolation between the
two adjacent data points. In a similar fashion SF was calculated for a large
number of IDV light curves (Sagar et al. 2004, Stalin et al. 2005). The behavior 
of the first order SF have the following types:

(i) If the source first order SF does not display any plateau, it represents
the time scale of the variability to be larger than the length of the data.

(ii) If the source first order SF display a plateau followed by a dip
in the SF, the dip indicates a possible cycle and plateau the 
time scale of the variability.

Fig. 3 displays the SF of the light curve of top panel of Fig 2. (g). SF 
display one plateau followed by a dip. From the plateau we deduce the variability 
time scales to be 0.01 day i.e $\sim$ 15 min. 

\section {Discussions \& Conclusions}

From our observations on Mrk 501 during March - May, 2000, we find the existence 
of IDV (to the extent of $\sim$ 6-7 \%) in R band on one occasion. 
The source has shown 
short term flux variability and total variation detected in our observations in V 
passband is $\sim$ 15 \% and in R passband is $\sim$ 25 \%. The data do not show 
any significant variation in (B$-$R) color but (V$-$R) color has varied 
by about 0.2 mag. Though we see some variation in brightness in the light curves, 
it appears that Mrk 501 was not in active phase during our observations.

We have detected variation, on April 5,  in time scale of $\sim$ 15 min in Mrk
501. 
Our observations do not show any periodicity or quasi-periodicity in intra-day or short
time scales. But, in some other blazars quasi periodic variations have been reported 
earlier on longer as well as shorter time scales e.g. quasi periodic variations,
 existing 
simultaneously in optical and radio wavelength, are reported on the timescale of 1 to 7 
days in S5 0716+714 (Quirrenbach et al. 1991). The same source on other occasion shows
quasi periodic variation in time scale of 4 days in optical region (Heidt \&
Wagner 1996). Quasi periodic variations  were also found in the PKS 2155-304 in
UV and X-ray wavelength (Urry et al. 1993). On short time scale, for
example,    
periodic flux variation is reported in OJ 287 in radio as well as in
optical  band. Valtaoja et al. (1985) have reported 15.7 min periodicity at
37 GHz, Carrasco et al. (1985) have reported 23 minute period in
their observations in optical B passband 
and Carini et al. (1992) have shown the optical periodicity of 32 min.

Several models have been developed to explain the IDV and short term
variability  in blazars viz. the shock-in-jet models, accretion-disk based
models (e.g. Wagner  \& Witzel 1995; Urry \& Padovani 1995; Ulrich, Maraschi \&
Urry 1997 and references  therein). The detection of quasi-periodicity in
time scale  of minutes to less than a week in some of the blazars could be
taken as evidence for the existence of accretion disk pulsation, or  the
presence of a single dominating hot-spot on the accretion disk,   or of a
helical  structure in a jet (Vila 1979, Mangalam \& Wiita 1993, Chakrabarti \&
Wiita 1993).  Quasi-periodic variation may also be attributed to a precessing
jet.  So search for periodic variations has great interests. The  variability
reported in  the present work could be explained by any of the above models. 

However, the variability detected in the low state of blazars can be
explained on the basis of some kind of instability in the 
accretion disk, because in the low state, jet emission is less
dominant over the thermal  emission from the disk. Within unified schemes, the
blazars are seen nearly face on, hence any accretion disk  fluctuations
should be directly visible. The models for the flares randomly emerging at
different locations and at different time on accretion disks could be very
relevant to the variations seen in blazars at low-state. Eclipse of the
hot spot by
parts of the disk between the individual spot and the observer could be 
the other possible
mechanism to produce variability in blazars at low-state. There
often exists a range of azimuthal positions at a given  radius within which the
spot is eclipsed, and this is directly depends on the  geometry of disk and
the viewing angle of observer.
The central black hole mass is expected to shed some light on the evolution process 
in AGNs (Barth et al. 2002, Fan 2003). There are several ways to find
the BH mass but there is no consensus on this and the BH mass reported in the
literature for a particular source differ significantly. The BH mass for Mrk 501, 
estimated by many authors, is in the range (0.01 $-$ 3.4) $\times$ 
10$^{9}$ M$_{\odot}$ (Barth et al. 2002, Fan et al. 1999, Merritt \& Ferrarese 2001, 
Rieger \& Mannheim 2003).
For Mrk 501, our optical observation show the variability time scale of 15
min. We assume that detected IDV is due to an instability in the vicinity of
the super massive BH and the size of emitting region is taken as R $=$ ct,
where t is the variability time scale (i.e 15min). Further assuming that the
15 min variability time scale corresponds to instability or disk pulsation on
the accretion disk at a distance of R = 5 R$_{S}$ where R$_{S} =$2GM/c$^{2}$
is Schwarschild radius, the  mass of the super massive BH, M can
be estimated by using the formula:  M = c$^{3}$t/10G = 1.20 $\times$ 10$^{8}$
M$_{\odot}$. The Eddington luminosity for a black  hole of mass M is given by
L$_{E} =$ 1.3 $\times$ 10$^{38}$  (M/M$_{\odot}$) erg s$^{-1}$ (Witta 1985),
which in the present case will be  L$_{E} =$ 1.56 $\times$ 10$^{46}$ erg
s$^{-1}$. The result is in the range of  (0.01 $-$ 3.4) $\times 10^{9}
M_{\odot}$, and hence  consistent with other results. 

There is an alternative way to explain the IDV detected in the present work.
In the recent work (e.g. Chiaberge et al. 1999, 2006; Chiaberge \& Macchetto 2007; 
Capetti et al. 2007; Macchetto \& Chiaberge 2007 and references therein), by analyzing 
R band HST images of a sample of 33 FR I sources, they were able to detect the 
unresolved optical nuclear sources. They found that optical and radio nuclear 
luminosities were strongly correlated. They also found tight correlation between the radio and
X-ray luminosities measured from Chandra observations (Balmaverde et al. 2006). Furthermore, 
the correlation between radio, optical and X-ray emission extends to the large population of
radio-loud AGNs found in early-type galaxies, with luminosities smaller by
a factor $\sim$ 1000 with respect to classical low-luminosity radio-galaxies
(Balmaverde \& Capetti 2006). According to these findings the nuclear emission 
in low luminosity radio-loud AGNs is dominated by non-thermal emission from the 
base of the jet. Assuming FR I and BL Lacs unification, which is yet to be established,
the alternate possibility is that the IDV seen in BL Lacs could arise in
the jet. Considering this model, the detected small IDV in the low-state
of the BL Lac object Mrk 501 might be due to a weak emission from the base
of the jet.

It will be really interesting to start a long term multi-color optical monitoring
program of BL Lac objects in their low-states to search for the IDV. We believe,
with focused effort, we can make a large data set, and we may be able to find 
which model is more appropriate to discuss the results of IDV in the low-state of 
BL Lac objects. At this stage, we leave it as an open problem.

We thank the anonymous referees for their useful suggestions. A. C. Gupta, 
W. G. Deng and J. M. Bai gratefully acknowledge the financial support from the 
National Natural Science Foundation of China (grant nos. 10533050 and 10573030). 
The research work at Physical Research Laboratory was funded by the Department 
of Space, Government of India. M. G. Lee acknowledges the financial support 
from the BK21 Project of the Korean government. IRAF is distributed by NOAO, USA.

\clearpage

\begin{figure}
\plotone{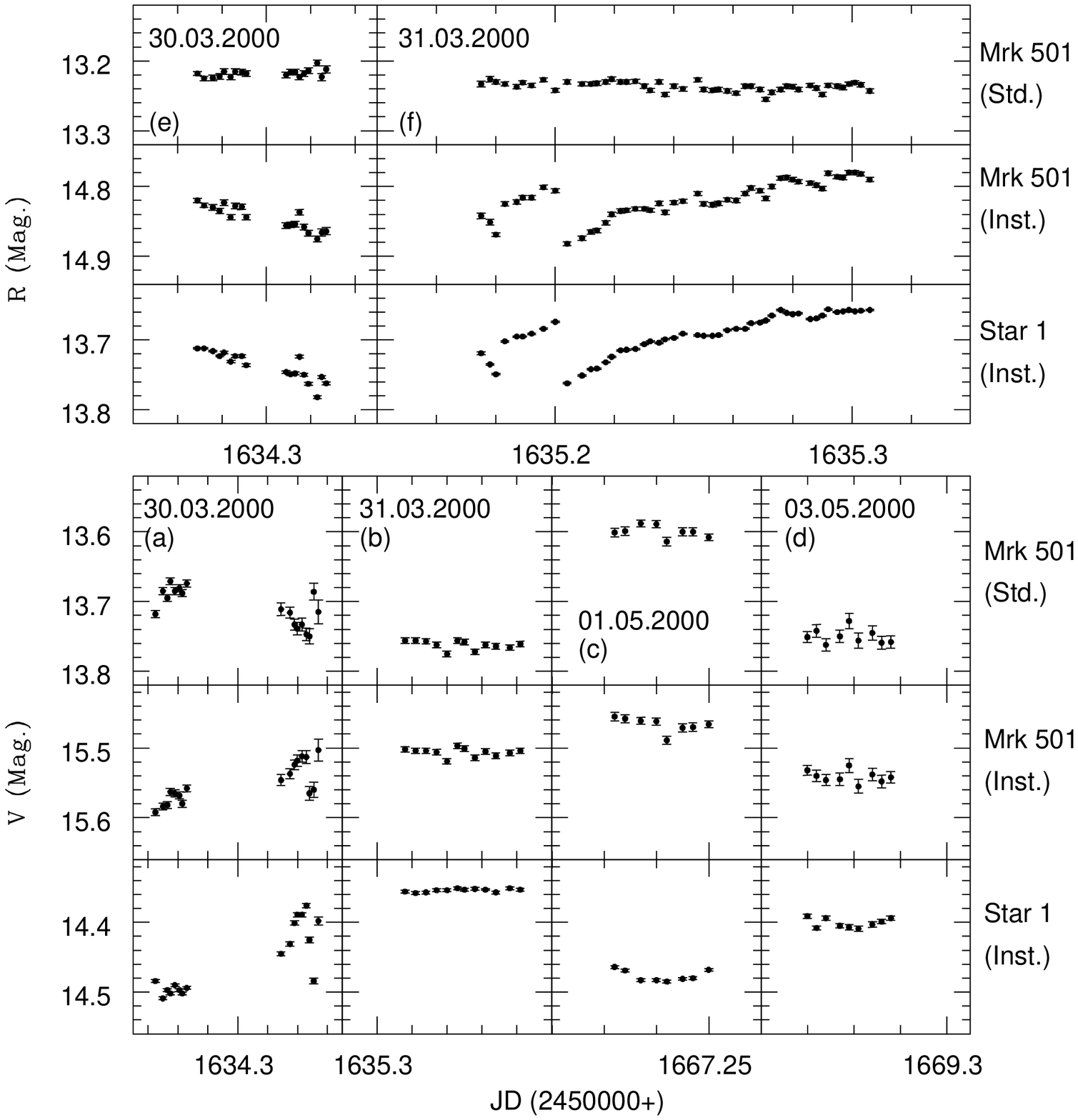}
\caption{Light-curves for V-band are plotted in panel (a), (b), (c) and (d) 
respectively for March 30, March 31, May 1 and May 3. The R-band light-curves  
for March 30 and March 31 are plotted  respectively in panel (e) and (f).}
\end{figure}

\begin{figure}
\plotone{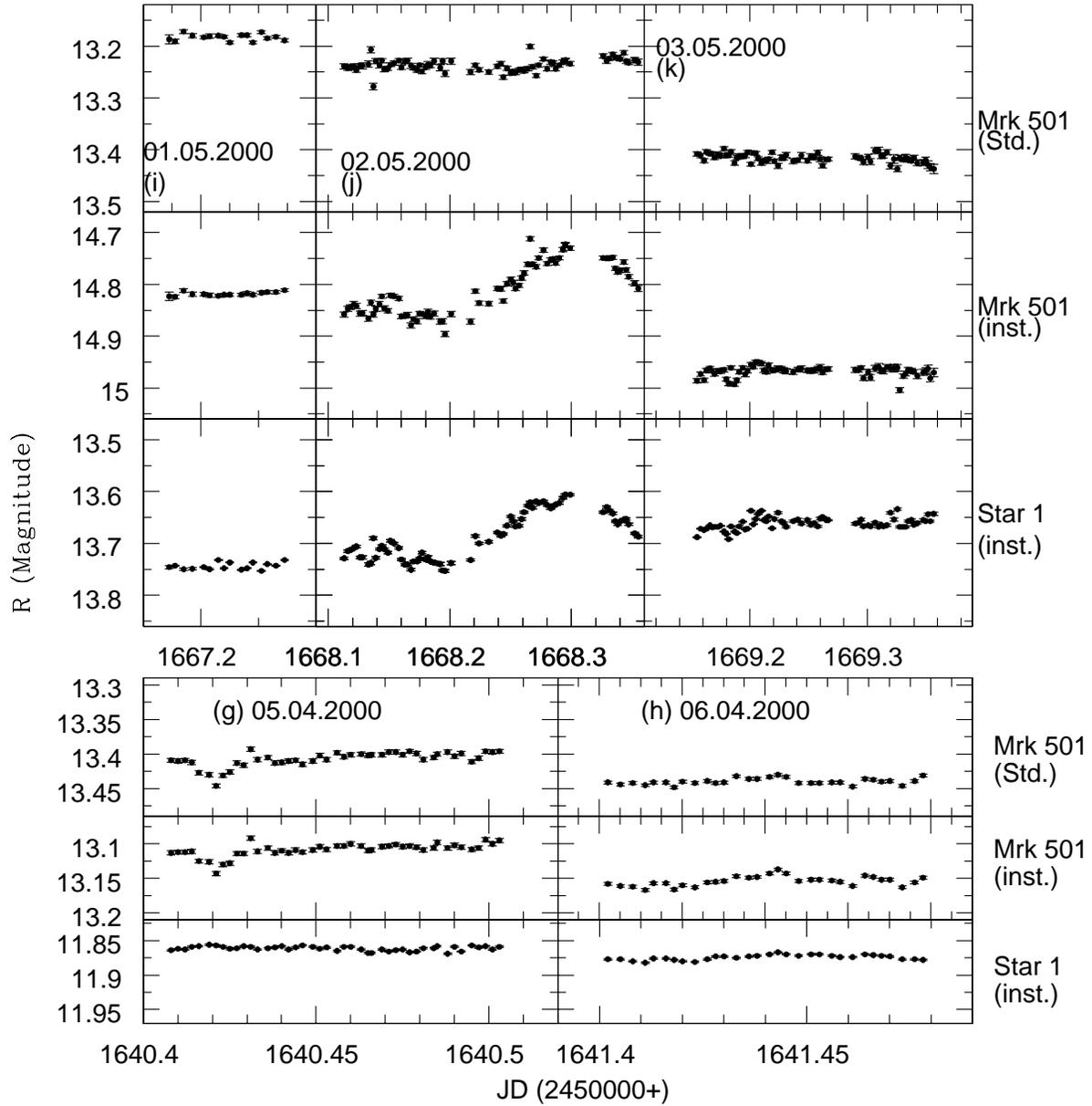}
\caption{The R-band light-curves  
for April 5, April 6, May 1, May 2 and May 3 
are plotted  respectively in panel (g), (h), (i), (j) and (k).}
\end{figure}

\begin{figure}
\plotone{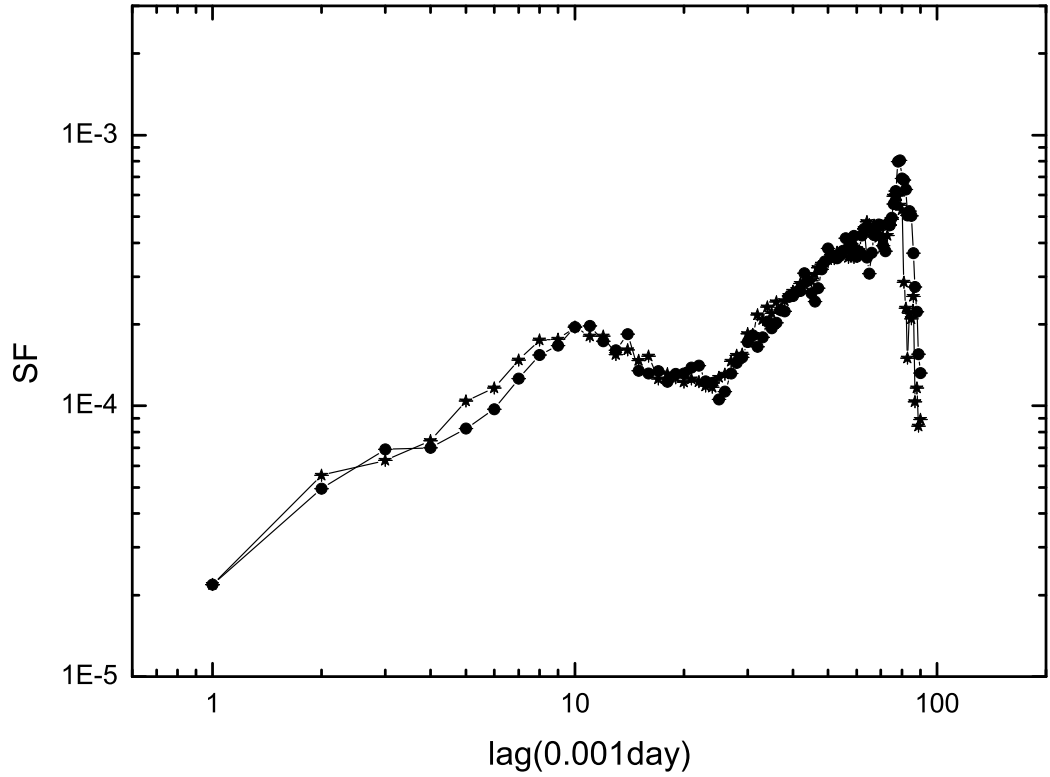}
\caption{First order structure function of the BL Lac object Mrk 501 of the top
panel of Fig. 2 (g). Error bars are also plotted in the figure but the size of error
bars are less than the size of the symbol, so, not visible clearly.} 
\end{figure}

\begin{figure}
\plotone{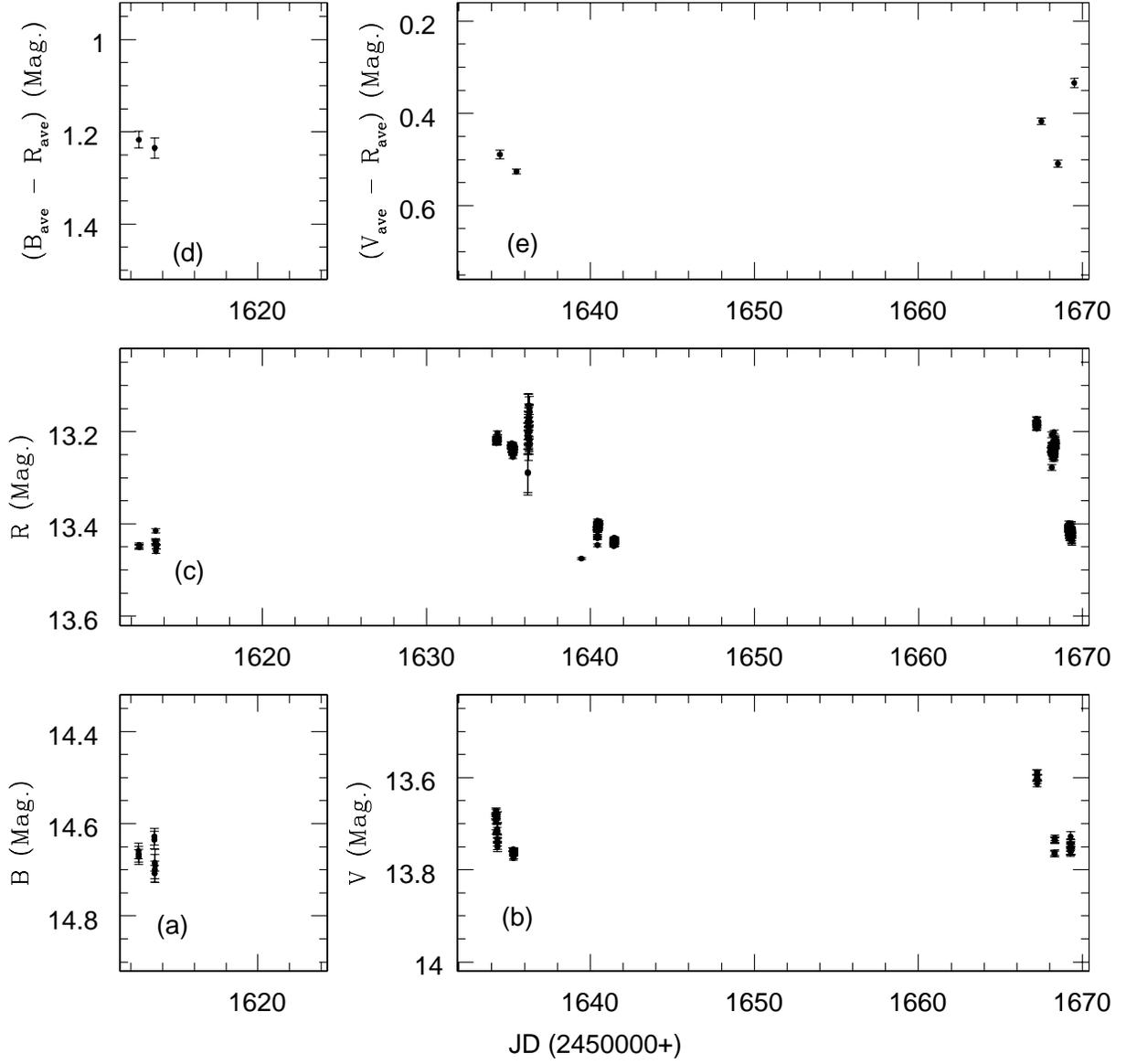}
\caption{Panels a, b, c, d and e represent light-curves 
in magnitude B,V, R  and color (B$_{ave}$ - R$_{ave}$
 and (V$_{ave}$ - R$_{ave}$) respectively for the 
 observing run March 8 - May 3, 2000.}
\end{figure}

\clearpage

\begin{table}
\caption[]{The complete log of observations of Mrk 501 and IDV status. V and NV 
represent intraday variability observed and not observed respectively.}
\begin{center}
\begin{tabular}{ccclcl} \hline \hline 
 Date       & Filter  & No. of  & Telescope & IDV      & Time \\
(dd.mm.yyyy)&         & data    &           & Status   & Scale \\
            &         & points  &           &          &  \\ \hline
 08.03.2000 & B       & ~3      &  India    &  NV      &        \\
            & R       & ~3      &  India    &  NV      &        \\
 09.03.2000 & B       & ~3      &  India    &  NV      &        \\
            & R       & ~4      &  India    &  NV      &        \\
 10.03.2000 & B       & ~3      &  India    &  NV      &        \\
            & R       & ~2      &  India    &  NV      &        \\
 30.03.2000 & V       & 17      &  Korea    &  NV      &        \\
            & R       & 18      &  Korea    &  NV      &        \\
 31.03.2000 & V       & 12      &  Korea    &  NV      &        \\
            & R       & 49      &  Korea    &  NV      &        \\
 01.04.2000 & R       & 32      &  Korea    &  Noisy   &        \\
 04.04.2000 & R       & ~1      &  India    &  NV      &        \\
 05.04.2000 & R       & 45      &  India    &  ~V      &  15 min  \\
 06.04.2000 & R       & 32      &  India    &  NV      &        \\
 01.05.2000 & V       & ~8      &  Korea    &  NV      &        \\
            & R       & 16      &  Korea    &  NV      &        \\
 02.05.2000 & V       & ~6      &  Korea    &  NV      &        \\
            & R       & 76      &  Korea    &  NV      &        \\
 03.05.2000 & V       & ~9      &  Korea    &  NV      &        \\
            & R       & 70      &  Korea    &  NV      &        \\
\hline
\end{tabular}
\end{center}
\end{table}

\begin{table}
\caption[]{The B-band data of Mrk 501.} 
\begin{center}
\begin{tabular}{ccc} \hline \hline 
 JD (2450000+)    &  Magnitude & Error  \\\hline
1612.479  & 14.659   &  0.017 \\
1612.486  & 14.666   &  0.017 \\
1612.493  & 14.672   &  0.016 \\
1613.478  & 14.635   &  0.019 \\
1613.487  & 14.628   &  0.018 \\
1613.494  & 14.708   &  0.018 \\
1613.501  & 14.685   &  0.018 \\
1613.516  & 14.700   &  0.019 \\
1613.525  & 14.691   &  0.036 \\
\hline
\end{tabular}
\end{center}
\end{table}

\begin{table}
\caption[]{The V-band data of Mrk 501.} 
\begin{center}
\begin{tabular}{cccccc} \hline \hline 
 JD (2450000+)    &  Magnitude & Error &  JD (2450000+)    &  Magnitude & Error \\\hline
 1634.245 &  13.718 &   0.005  & 1635.334 &  13.764 &   0.004 \\
 1634.250 &  13.685 &   0.005  & 1635.338 &  13.766 &   0.004 \\
 1634.253 &  13.695 &   0.005  & 1635.341 &  13.761 &   0.004 \\
 1634.255 &  13.671 &   0.005  & 1667.232 &  13.601 &   0.006 \\
 1634.258 &  13.685 &   0.005  & 1667.234 &  13.599 &   0.006 \\
 1634.261 &  13.681 &   0.005  & 1667.237 &  13.588 &   0.005 \\
 1634.263 &  13.688 &   0.005  & 1667.240 &  13.589 &   0.005 \\
 1634.266 &  13.674 &   0.005  & 1667.242 &  13.614 &   0.006 \\
 1634.329 &  13.711 &   0.009  & 1667.245 &  13.600 &   0.006  \\
 1634.335 &  13.716 &   0.008  & 1667.247 &  13.600 &   0.006 \\
 1634.338 &  13.733 &   0.008  & 1667.250 &  13.608 &   0.005  \\
 1634.340 &  13.739 &   0.009  & 1668.303 &  13.763 &   0.006 \\
 1634.343 &  13.733 &   0.009  & 1668.307 &  13.766 &   0.006  \\
 1634.346 &  13.747 &   0.009  & 1668.309 &  13.735 &   0.006  \\
 1634.348 &  13.750 &   0.011  & 1668.317 &  13.731 &   0.006  \\
 1634.351 &  13.686 &   0.012  & 1668.321 &  13.735 &   0.006  \\
 1634.354 &  13.715 &   0.017  & 1668.323 &  13.737 &   0.006 \\
 1635.308 &  13.756 &   0.004  & 1669.270 &  13.751 &   0.008  \\
 1635.311 &  13.756 &   0.004  & 1669.272 &  13.742 &   0.009 \\
 1635.314 &  13.757 &   0.004  & 1669.274 &  13.762 &   0.009 \\
 1635.317 &  13.762 &   0.004  & 1669.277 &  13.750 &   0.009 \\
 1635.320 &  13.775 &   0.004  & 1669.279 &  13.728 &   0.011 \\
 1635.323 &  13.756 &   0.004  & 1669.281 &  13.756 &   0.011 \\
 1635.325 &  13.758 &   0.004  & 1669.284 &  13.745 &   0.010 \\
 1635.328 &  13.772 &   0.004  & 1669.286 &  13.759 &   0.009 \\
 1635.331 &  13.762 &   0.004  & 1669.288 &  13.758 &   0.009 \\
\hline
\end{tabular}
\end{center}
\end{table}

\begin{table}
\caption[]{The R-band data of Mrk 501.} 
\scriptsize
\begin{center}
\begin{tabular}{ccccccccc} \hline \hline 
 JD (2450000+)    &  Magnitude & Error & JD (2450000+)&  Magnitude & Error &  JD (2450000+) &  Magnitude & Error  \\\hline
 1612.474  &   13.446 &   0.005 &   1635.217 &    13.230  &  0.003 &   1636.184 &    13.290 &   0.042 \\
 1612.483  &   13.450 &   0.005 &   1635.219 &    13.226  &  0.003 &   1636.186 &    13.215 &   0.035 \\
 1612.490  &   13.450 &   0.005 &   1635.222 &    13.230  &  0.003 &   1636.203 &    13.212 &   0.016 \\
 1613.475  &   13.437 &   0.005 &   1635.224 &    13.230  &  0.003 &   1636.206 &    13.199 &   0.010 \\
 1613.483  &   13.415 &   0.005 &   1635.227 &    13.229  &  0.003 &   1636.208 &    13.180 &   0.010 \\
 1613.491  &   13.442 &   0.005 &   1635.230 &    13.236  &  0.003 &   1636.211 &    13.176 &   0.013 \\
 1613.498  &   13.439 &   0.005 &   1635.232 &    13.242  &  0.003 &   1636.214 &    13.203 &   0.014 \\
 1613.505  &   13.459 &   0.005 &   1635.235 &    13.230  &  0.003 &   1636.216 &    13.199 &   0.029 \\
 1613.516  &   13.450 &   0.005 &   1635.237 &    13.248  &  0.003 &   1636.218 &    13.177 &   0.022 \\
 1634.269  &   13.218 &   0.003 &   1635.240 &    13.236  &  0.003 &   1636.220 &    13.176 &   0.019 \\
 1634.272  &   13.225 &   0.003 &   1635.243 &    13.240  &  0.003 &   1636.222 &    13.214 &   0.018 \\
 1634.276  &   13.224 &   0.004 &   1635.248 &    13.227  &  0.003 &   1636.224 &    13.190 &   0.033 \\
 1634.279  &   13.222 &   0.004 &   1635.250 &    13.241  &  0.003 &   1636.226 &    13.232 &   0.030 \\
 1634.281  &   13.215 &   0.004 &   1635.253 &    13.242  &  0.003 &   1636.228 &    13.189 &   0.040 \\
 1634.284  &   13.223 &   0.004 &   1635.255 &    13.241  &  0.003 &   1636.240 &    13.143 &   0.025 \\
 1634.286  &   13.215 &   0.004 &   1635.258 &    13.243  &  0.003 &   1636.242 &    13.214 &   0.023 \\
 1634.289  &   13.216 &   0.004 &   1635.261 &    13.246  &  0.003 &   1636.245 &    13.165 &   0.025 \\
 1634.291  &   13.218 &   0.004 &   1635.264 &    13.236  &  0.003 &   1636.248 &    13.172 &   0.026 \\
 1634.309  &   13.220 &   0.004 &   1635.266 &    13.236  &  0.003 &   1636.250 &    13.212 &   0.032 \\
 1634.311  &   13.216 &   0.004 &   1635.269 &    13.241  &  0.003 &   1636.252 &    13.179 &   0.036 \\
 1634.313  &   13.216 &   0.004 &   1635.271 &    13.255  &  0.003 &   1636.254 &    13.202 &   0.024 \\
 1634.315  &   13.223 &   0.004 &   1635.273 &    13.245  &  0.003 &   1636.256 &    13.219 &   0.022 \\
 1634.317  &   13.218 &   0.004 &   1635.276 &    13.241  &  0.003 &   1636.258 &    13.160 &   0.018 \\
 1634.319  &   13.214 &   0.004 &   1635.278 &    13.236  &  0.003 &   1636.260 &    13.198 &   0.013 \\
 1634.323  &   13.203 &   0.004 &   1635.280 &    13.237  &  0.003 &   1636.262 &    13.175 &   0.012 \\
 1634.325  &   13.223 &   0.005 &   1635.282 &    13.241  &  0.003 &   1636.278 &    13.151 &   0.027 \\
 1634.327  &   13.212 &   0.005 &   1635.286 &    13.235  &  0.003 &   1636.282 &    13.179 &   0.037 \\
 1635.175  &   13.233 &   0.004 &   1635.288 &    13.239  &  0.003 &   1636.284 &    13.218 &   0.031 \\
 1635.178  &   13.226 &   0.004 &   1635.290 &    13.248  &  0.003 &   1639.445 &    13.475 &   0.002 \\
 1635.180  &   13.230 &   0.003 &   1635.292 &    13.235  &  0.003 &   1640.408 &    13.409 &   0.003 \\
 1635.183  &   13.233 &   0.003 &   1635.295 &    13.236  &  0.003 &   1640.410 &    13.410 &   0.003 \\
 1635.187  &   13.237 &   0.003 &   1635.297 &    13.238  &  0.003 &   1640.412 &    13.409 &   0.003 \\
 1635.189  &   13.231 &   0.003 &   1635.299 &    13.233  &  0.003 &   1640.414 &    13.412 &   0.003 \\
 1635.192  &   13.235 &   0.003 &   1635.301 &    13.231  &  0.003 &   1640.416 &    13.427 &   0.003 \\
 1635.196  &   13.227 &   0.003 &   1635.303 &    13.234  &  0.003 &   1640.419 &    13.430 &   0.003 \\
 1635.200  &   13.242 &   0.003 &   1635.306 &    13.243  &  0.003 &   1640.421 &    13.446 &   0.003 \\
 1635.204  &   13.230 &   0.003 &   1636.176 &    13.207  &  0.010 &   1640.423 &    13.431 &   0.003 \\
 1635.209  &   13.233 &   0.003 &   1636.178 &    13.204  &  0.022 &   1640.425 &    13.426 &   0.003 \\
 1635.212  &   13.233 &   0.003 &   1636.180 &    13.288  &  0.049 &   1640.427 &    13.413 &   0.003 \\
 1635.214  &   13.232 &   0.003 &   1636.182 &    13.178  &  0.060 &   1640.429 &    13.416 &   0.003 \\
\hline 
\end{tabular}
\end{center}
\end{table}

\begin{table}
\scriptsize
\begin{center}
\begin{tabular}{ccccccccc} \hline \hline 
 JD (2450000+)    &  Magnitude & Error & JD (2450000+)&  Magnitude & Error &  JD (2450000+) &  Magnitude & Error  \\\hline
 1640.431  &   13.393 &   0.003 &   1641.418  &   13.448 &   0.002 &   1667.229  &   13.189 &   0.003 \\
 1640.433  &   13.408 &   0.003 &   1641.420  &   13.440 &   0.002 &   1668.113  &   13.239 &   0.005 \\
 1640.436  &   13.405 &   0.003 &   1641.423  &   13.442 &   0.002 &   1668.116  &   13.242 &   0.005 \\
 1640.438  &   13.413 &   0.003 &   1641.426  &   13.439 &   0.002 &   1668.118  &   13.241 &   0.005 \\
 1640.440  &   13.412 &   0.003 &   1641.428  &   13.442 &   0.002 &   1668.121  &   13.239 &   0.005 \\
 1640.442  &   13.410 &   0.003 &   1641.430  &   13.441 &   0.002 &   1668.123  &   13.246 &   0.005 \\
 1640.444  &   13.409 &   0.003 &   1641.433  &   13.432 &   0.002 &   1668.126  &   13.239 &   0.005 \\
 1640.446  &   13.415 &   0.003 &   1641.436  &   13.436 &   0.002 &   1668.128  &   13.238 &   0.006 \\
 1640.449  &   13.410 &   0.003 &   1641.438  &   13.436 &   0.002 &   1668.133  &   13.235 &   0.006 \\
 1640.451  &   13.402 &   0.003 &   1641.441  &   13.433 &   0.002 &   1668.135  &   13.207 &   0.006 \\
 1640.453  &   13.408 &   0.003 &   1641.443  &   13.430 &   0.002 &   1668.137  &   13.278 &   0.006 \\
 1640.456  &   13.398 &   0.003 &   1641.445  &   13.433 &   0.002 &   1668.139  &   13.229 &   0.005 \\
 1640.458  &   13.404 &   0.003 &   1641.448  &   13.442 &   0.002 &   1668.142  &   13.237 &   0.005 \\
 1640.460  &   13.401 &   0.003 &   1641.451  &   13.442 &   0.002 &   1668.144  &   13.228 &   0.004 \\
 1640.463  &   13.400 &   0.003 &   1641.453  &   13.442 &   0.002 &   1668.146  &   13.245 &   0.005 \\
 1640.465  &   13.402 &   0.003 &   1641.456  &   13.441 &   0.002 &   1668.149  &   13.243 &   0.004 \\
 1640.466  &   13.401 &   0.003 &   1641.458  &   13.441 &   0.002 &   1668.151  &   13.236 &   0.004 \\
 1640.469  &   13.401 &   0.003 &   1641.461  &   13.447 &   0.002 &   1668.154  &   13.233 &   0.004 \\
 1640.471  &   13.397 &   0.003 &   1641.464  &   13.436 &   0.002 &   1668.158  &   13.228 &   0.004 \\
 1640.473  &   13.397 &   0.003 &   1641.466  &   13.437 &   0.002 &   1668.160  &   13.241 &   0.004 \\
 1640.475  &   13.401 &   0.003 &   1641.468  &   13.440 &   0.002 &   1668.163  &   13.230 &   0.004 \\
 1640.477  &   13.396 &   0.003 &   1641.470  &   13.439 &   0.002 &   1668.165  &   13.228 &   0.005 \\
 1640.479  &   13.399 &   0.003 &   1641.473  &   13.446 &   0.002 &   1668.168  &   13.238 &   0.005 \\
 1640.481  &   13.408 &   0.003 &   1641.476  &   13.439 &   0.002 &   1668.170  &   13.242 &   0.004 \\
 1640.484  &   13.405 &   0.003 &   1641.478  &   13.431 &   0.002 &   1668.173  &   13.246 &   0.005 \\
 1640.485  &   13.400 &   0.003 &   1667.189  &   13.187 &   0.009 &   1668.175  &   13.237 &   0.004 \\
 1640.488  &   13.397 &   0.003 &   1667.191  &   13.191 &   0.004 &   1668.177  &   13.249 &   0.004 \\
 1640.490  &   13.403 &   0.003 &   1667.194  &   13.172 &   0.004 &   1668.180  &   13.240 &   0.005 \\
 1640.492  &   13.399 &   0.003 &   1667.197  &   13.180 &   0.004 &   1668.182  &   13.236 &   0.006 \\
 1640.495  &   13.411 &   0.003 &   1667.201  &   13.183 &   0.003 &   1668.184  &   13.237 &   0.005 \\
 1640.497  &   13.406 &   0.003 &   1667.203  &   13.181 &   0.004 &   1668.187  &   13.229 &   0.006 \\
 1640.499  &   13.396 &   0.003 &   1667.206  &   13.180 &   0.003 &   1668.192  &   13.242 &   0.006 \\
 1640.501  &   13.397 &   0.003 &   1667.208  &   13.182 &   0.003 &   1668.194  &   13.229 &   0.005 \\
 1640.503  &   13.396 &   0.003 &   1667.210  &   13.193 &   0.003 &   1668.196  &   13.253 &   0.006 \\
 1641.402  &   13.441 &   0.002 &   1667.214  &   13.179 &   0.003 &   1668.201  &   13.229 &   0.006 \\
 1641.405  &   13.444 &   0.002 &   1667.216  &   13.179 &   0.003 &   1668.217  &   13.250 &   0.005 \\
 1641.408  &   13.442 &   0.002 &   1667.218  &   13.193 &   0.003 &   1668.221  &   13.237 &   0.004 \\
 1641.411  &   13.445 &   0.002 &   1667.221  &   13.173 &   0.003 &   1668.224  &   13.246 &   0.004 \\
 1641.413  &   13.441 &   0.002 &   1667.223  &   13.185 &   0.003 &   1668.232  &   13.250 &   0.004 \\
 1641.416  &   13.441 &   0.002 &   1667.226  &   13.182 &   0.003 &   1668.239  &   13.239 &   0.004 \\
\hline 
\end{tabular}
\end{center}
\end{table}

\begin{table}
\scriptsize
\begin{center}
\begin{tabular}{ccccccccc} \hline \hline 
 JD (2450000+)    &  Magnitude & Error & JD (2450000+)&  Magnitude & Error &  JD (2450000+) &  Magnitude & Error  \\\hline
 1668.242  &   13.234 &   0.004  &  1669.155  &   13.408 &   0.004 &   1669.244   &  13.411 &   0.006  \\
 1668.244  &   13.260 &   0.004  &  1669.158  &   13.411 &   0.004 &   1669.249   &  13.421 &   0.004  \\
 1668.247  &   13.243 &   0.004  &  1669.161  &   13.421 &   0.004 &   1669.252   &  13.413 &   0.005  \\
 1668.250  &   13.252 &   0.004  &  1669.163  &   13.404 &   0.004 &   1669.255   &  13.415 &   0.005  \\
 1668.252  &   13.249 &   0.004  &  1669.166  &   13.407 &   0.004 &   1669.258   &  13.406 &   0.005  \\
 1668.254  &   13.251 &   0.004  &  1669.168  &   13.408 &   0.004 &   1669.260   &  13.418 &   0.005  \\
 1668.257  &   13.246 &   0.004  &  1669.170  &   13.415 &   0.004 &   1669.262   &  13.431 &   0.004  \\
 1668.259  &   13.245 &   0.004  &  1669.173  &   13.409 &   0.004 &   1669.265   &  13.419 &   0.004  \\
 1668.261  &   13.248 &   0.004  &  1669.175  &   13.412 &   0.004 &   1669.267   &  13.418 &   0.004  \\
 1668.264  &   13.244 &   0.004  &  1669.178  &   13.398 &   0.004 &   1669.290   &  13.413 &   0.005  \\
 1668.266  &   13.201 &   0.004  &  1669.180  &   13.411 &   0.004 &   1669.293   &  13.415 &   0.004  \\
 1668.268  &   13.240 &   0.004  &  1669.182  &   13.410 &   0.004 &   1669.295   &  13.417 &   0.004  \\
 1668.271  &   13.257 &   0.004  &  1669.184  &   13.404 &   0.004 &   1669.297   &  13.424 &   0.005  \\
 1668.273  &   13.237 &   0.004  &  1669.187  &   13.426 &   0.004 &   1669.301   &  13.414 &   0.005  \\
 1668.277  &   13.225 &   0.004  &  1669.189  &   13.415 &   0.004 &   1669.303   &  13.423 &   0.006  \\
 1668.280  &   13.244 &   0.004  &  1669.191  &   13.412 &   0.004 &   1669.307   &  13.402 &   0.006  \\
 1668.285  &   13.233 &   0.004  &  1669.194  &   13.413 &   0.004 &   1669.310   &  13.401 &   0.006  \\
 1668.283  &   13.231 &   0.004  &  1669.196  &   13.411 &   0.004 &   1669.312   &  13.412 &   0.005  \\
 1668.287  &   13.244 &   0.004  &  1669.198  &   13.405 &   0.004 &   1669.314   &  13.411 &   0.005  \\
 1668.290  &   13.237 &   0.004  &  1669.201  &   13.428 &   0.004 &   1669.317   &  13.405 &   0.005  \\
 1668.293  &   13.231 &   0.004  &  1669.203  &   13.406 &   0.004 &   1669.320   &  13.431 &   0.006  \\
 1668.295  &   13.229 &   0.004  &  1669.205  &   13.407 &   0.004 &   1669.323   &  13.418 &   0.008  \\
 1668.297  &   13.227 &   0.004  &  1669.207  &   13.417 &   0.004 &   1669.326   &  13.437 &   0.006  \\
 1668.299  &   13.234 &   0.004  &  1669.210  &   13.426 &   0.004 &   1669.328   &  13.416 &   0.005  \\
 1668.326  &   13.219 &   0.004  &  1669.212  &   13.421 &   0.004 &   1669.331   &  13.419 &   0.004  \\
 1668.329  &   13.229 &   0.004  &  1669.214  &   13.422 &   0.004 &   1669.333   &  13.414 &   0.005  \\
 1668.331  &   13.224 &   0.004  &  1669.216  &   13.418 &   0.004 &   1669.335   &  13.418 &   0.005  \\
 1668.334  &   13.216 &   0.004  &  1669.219  &   13.405 &   0.004 &   1669.338   &  13.421 &   0.004  \\
 1668.336  &   13.222 &   0.004  &  1669.221  &   13.422 &   0.004 &   1669.340   &  13.415 &   0.004  \\
 1668.338  &   13.223 &   0.004  &  1669.224  &   13.431 &   0.005 &   1669.343   &  13.426 &   0.005  \\
 1668.340  &   13.225 &   0.004  &  1669.227  &   13.416 &   0.005 &   1669.348   &  13.425 &   0.006  \\
 1668.343  &   13.213 &   0.004  &  1669.229  &   13.415 &   0.005 &   1669.350   &  13.420 &   0.006  \\
 1668.345  &   13.229 &   0.004  &  1669.231  &   13.410 &   0.005 &   1669.352   &  13.429 &   0.007  \\
 1668.347  &   13.232 &   0.004  &  1669.237  &   13.422 &   0.005 &   1669.354   &  13.434 &   0.007  \\
 1668.352  &   13.227 &   0.005  &  1669.241  &   13.420 &   0.004 &   1669.357   &  13.437 &   0.009  \\
 1668.355  &   13.231 &   0.006  &            &          &         &              &         &  \\
\hline 
\end{tabular}
\end{center}
\end{table}

\end{document}